\documentclass[prl,aps,floats,twocolumn,showpacs,superscriptaddress,floatfix,amsmath]{revtex4}

\usepackage[a4paper,margin=2cm]{geometry}

\def\bc{\begin{center}}
\def\ec{\end{center}}
\def\beq{\begin{equation}}
\def\eeq{\end{equation}}
\def\bw{\begin{widetext}}
\def\ew{\end{widetext}}
\def\bea{\begin{eqnarray}}
\def\eea{\end{eqnarray}}

\usepackage{multirow}

\usepackage{graphicx}
\usepackage{dcolumn}
\usepackage{bm}

\begin{document}
\title{Higher Energy Composite Fermion Levels in the Fractional Quantum Hall Effect}

\author{Trevor D. Rhone}
\affiliation{Department of Physics, Columbia University, New York, NY 10027, USA}

\author{Dwipesh Majumder}
\affiliation{Department of Theoretical Physics, Indian Association for the Cultivation of Science, Jadavpur, Kolkata 700 032, India}

\author{Brian S. Dennis}
\affiliation{Department of Physics and Astronomy, Rutgers
University, Piscataway, NJ 08854, USA}

\author{Cyrus Hirjibehedin}
\affiliation{London Centre for Nanotechnology, University College
London, London WC1H OAH, United Kingdom}

\author{Irene Dujovne}
\affiliation{Department of Appl. Phys. and Appl. Math., Columbia
Univ., New York, NY 10027, USA}
\author{Javier G. Groshaus}
\affiliation{Institute for Optical Sciences, University of Toronto,
Toronto, Ontario, M5S 3H6, Canada}

\author{Yann Gallais}
\affiliation{Laboratoire Mat\'{e}riaux et Ph\'{e}nom\`{e}nes Quantiques, UMR CNRS 7162,
Universit\'{e} Paris Diderot 75205 Paris, France}

\author{Jainendra K. Jain}
\affiliation{Physics Department, Pennsylvania State University, University Park, PA 16802, USA}

\author{Sudhansu S. Mandal}
\affiliation{Department of Theoretical Physics, Indian Association for the Cultivation of Science, Jadavpur, Kolkata 700 032, India}

\author{Aron Pinczuk}
\affiliation{Department of Physics, Columbia University, New York,
NY 10027, USA} \affiliation{Department of Appl. Phys. and Appl.
Math., Columbia Univ., New York, NY 10027, USA}

\author{Loren Pfeiffer}
\author{Ken West}
\affiliation{Physics Department, Princeton University, Princeton, NJ 08544, USA}

\begin{abstract}
Even though composite fermions (CFs) in the fractional quantum Hall
liquid are well established, it is not yet known up to what energies
they remain intact. We probe the high-energy spectrum of
the 1/3 liquid directly by resonant inelastic light scattering (ILS), and
report the observation of a large number of new collective modes. Supported by our
theoretical calculations, we associate these with transitions across
two or more CF levels. Formation of quasiparticle levels up to high
energies is direct evidence for the robustness of topological order in the
fractional quantum Hall effect.
\end{abstract}

\date{\today}

\maketitle

\begin{figure}
  \centering
\includegraphics[width=2.9in,height=1.9in]{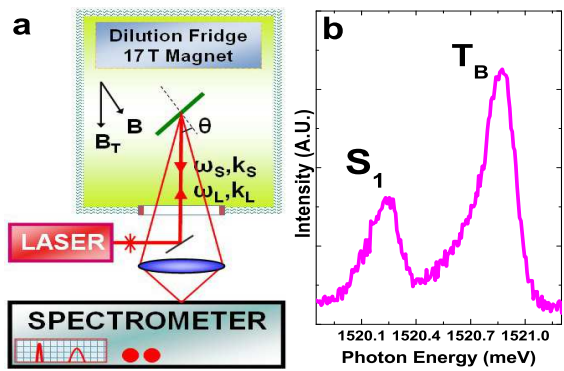}
\caption{(a) Schematic description of the light scattering set-up.
$\omega_L$ and $\omega_S$ are the frequencies of the incident laser photons (L)
and of the scattered photons (S).  (b) Optical emission at $\nu$=1/3 ($B_T$=8.0 T, 70 mK).
The two peaks, labeled $S_1$ and $T_B$ are two of the fundamental optical excitons of the GaAs quantum well;  tuning the incident photon energy close to one of these peaks produces a resonance enhancement of
ILS intensities.
} \label{setup}
\end{figure}

Collective states of matter have proved enormously important both because of
the conceptual structures they reveal and the role they play in
technological innovation. The fractional quantum Hall (FQH) liquid,
which emerges as a result of interactions between electrons when the
dimensionality is reduced to two and the Hilbert space is further
restricted by application of an intense magnetic field
\cite{Stormer99}, represents a cooperative behavior that does not
subscribe to concepts such as Bose-Einstein condensation, diagonal
or off-diagonal long range order, and Landau order parameter.
It is the realization of a topological quantum state of matter, the
understanding of which has influenced development in a wide variety
of fields, such as topological insulators, cold atoms, graphene,
generalized particle statistics, quantum cryptography, and more
\cite{Stern09,Simon08,Andrei09,Bolotin09,Callan,Spielman}.

Neutral excitations provide a window into the physics of the FQH liquid.
Early theoretical treatments of the lowest neutral collective mode of the
FQH state at $\nu$=1/3 employed a single mode approximation
\cite{Girvin85}, as well as exact diagonalization studies on small systems
\cite{haldane85}, and showed a minimum
in the dispersion, which, following the terminology used in superfluid Helium,
is called a ``magneto-roton."
Subsequently, the collective modes at this and other
fractions were
understood in terms of CFs, quasiparticles that
result from a binding of electrons and an even number of
quantized vortices \cite{Jain89}.
Despite their complex
collective character, CFs act as almost free
particles insofar as the low energy behavior is concerned
\cite{Stormer99}. They experience an effective magnetic field and
form their own Landau-like levels, which are called ``$\Lambda$ levels."
(The CF ``$\Lambda$ levels reside within the lowest electronic Landau level.) The neutral excitations are described as inter-$\Lambda$-level exciton collective modes of CFs \cite{Dev92,song94,halperin94,Scarola00},
in close analogy to the electronic collective modes of the integral Hall
states.

\begin{figure*}
\includegraphics[width=5.5in,height=3.9in]{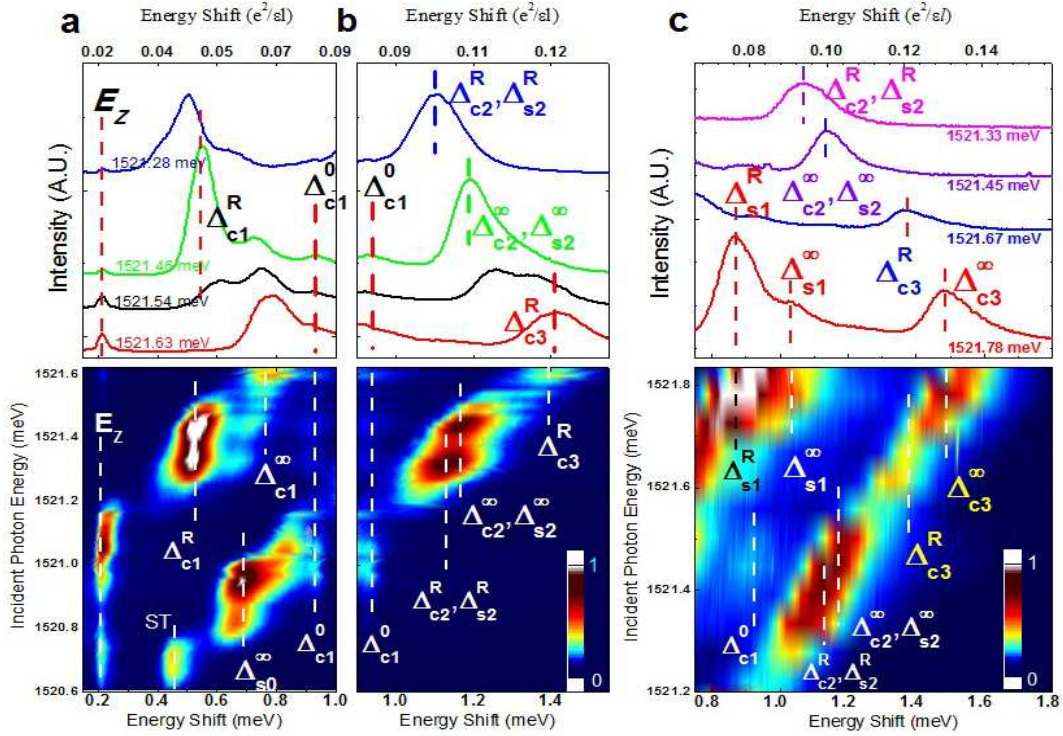}
\caption{ILS spectra of excitations
at $\nu$=1/3 as a function of the energy shift (with total magnetic
field $B_T$ $=$8.0 T, and a tilt of $30^o$). The energy is shown in units of
$e^2/\epsilon l$ on the top scale, where $l$ is magnetic length and $\epsilon$, the dielectric constant
of GaAs. The upper panels show peaks of several modes for certain
selected incident photon energies. The
lower panel contains a color plot of the
intensities of both (a) ``low energy" and (b),(c) the novel high energy modes. The vertical lines mark the positions of the
collective modes. The symbols, explained in the text, identify
the modes with excitations of CFs across several $\Lambda$ levels, both with and without spin reversal.} \label{1_3plot}
\end{figure*}

We report the excitation spectrum of the FQH fluid at $\nu$=1/3
in an unexplored energy range. Our main finding
is the existence of several well defined collective modes at energies
substantially exceeding those of the highest before reported
spin-conserving (SC) and spin-flip (SF) modes
\cite{Groshaus,Murthy09,Davies97}. Further, we provide compelling evidence,
supported by a detailed comparison between theory and experiment,
that these neutral modes represent a new family of excitations
involving CF transitions across several $\Lambda$ levels.
The direct experimental observation of the integrity
of $\Lambda$ levels at energies far above the Fermi energy demonstrates
that CFs are more robust than previously thought,
bolstering the expectation that the quasiparticles of
other topological states of CFs, such as the
nonabelian quasiparticles of the Pfaffian state at 5/2
\cite{Simon08,Stern10}, will also have comparably robust character.

The collective excitations of the FQH systems are measured by ILS. The
experiments are performed in a backscattering geometry with windows for
direct optical access to the sample, as shown in Fig. 1a.
The 2D electron system studied here
is formed in an asymmetrically doped, 33 nm wide
GaAs single quantum well (QW). The electron density
is n=5.6$\times$10$^{10}$cm$^{-2}$, with mobility,
$\mu$=7$\times$10$^6$cm$^2$Vs at T=300 mK.  Samples are mounted on the cold finger  of a dilution
refrigerator with a base temperature of 40 mK
that is inserted into the cold bore of a 17 T superconducting
magnet.  The energy of the incident photons,
$\omega_L$, is continuously tunable to be close to singlet ($S_1$) and triplet ($T_B$) fundamental optical transitions of the GaAs,
seen in emission spectra shown in Fig. \ref{setup}b \cite{Groshaus}.
The power density
is kept below 10$^{-4}$Wcm$^{-2}$.  Scattered light is dispersed by a Spex 1404 double
Czerny–Turner spectrometer with holographic master gratings.
Spectra are acquired by optical multichannel detection. The combined resolution of the system
is about 20$\mu$eV. Spectra can be taken with the linear
polarization of $\omega_L$ parallel (polarized)
or perpendicular (depolarized) to the detected scattered photon polarization.

The wave vector transferred from the photons to the 2D system is $q=(2\omega_L/c) \sin\theta$, much
smaller than $1/\emph{l}$, where l=$\sqrt{\hbar c/eB}$ is magnetic length.
However, weak short-range disorder induces a breakdown of
wave vector conservation \cite{Groshaus,Davies97,Marmorkos92,note2}, which
allows ILS to detect the critical points in
the exciton dispersion, such as the rotons,
because of van Hove singularities in the density of states at these energies.

The intensity of the ILS at $\nu$=1/3
is displayed in Fig. \ref{1_3plot} as a function of the energy transfer $\omega=\omega_L-\omega_S$. Each
peak indicates the presence of a collective mode. The collective mode energies
are marked by vertical lines\cite{comment4}.
The previously observed modes lie at
energies below $\sim$1 meV, as seen in Fig. \ref{1_3plot}a.
The striking feature of the spectra shown
in Fig.  \ref{1_3plot}b and \ref{1_3plot}c is the
existence of several new modes
up to 1.6 meV, the largest energy exchange
accessed in our experiments.

It is natural to interpret these new modes in terms of
excitations of CFs across K levels, referred to
below as ``level-K excitons."  Previous experiments at $\nu=1/3$ had reported only level-1 SC
excitons and level-0 SF excitons \cite{Groshaus,Davies97,Pinczuk93,Hirjibehedin05}.  Level-2 and level-3 CF excitons were recently investigated theoretically\cite{Majumder09} in the context of the splitting of the 1/3
collective mode at small but nonzero wave vectors
\cite{Hirjibehedin05}. Because the modes may also involve spin reversal, we adopt the notation in which we denote the level-$K$ spin-conserving modes by $\Delta_{cK}^\alpha$ and the level-$K$ spin-flip modes by $\Delta_{sK}^\alpha$. The superscript indicates the position of the mode: we have $\alpha=0$ for the zero wave vector mode, $\alpha=\infty$ for the large wave vector limit, and $\alpha=$R for a roton mode.  Identifications of the various modes shown on Fig.  \ref{1_3plot} are based on the analysis below.

\begin{figure}
\includegraphics[width=3.4in,height=3.7in]{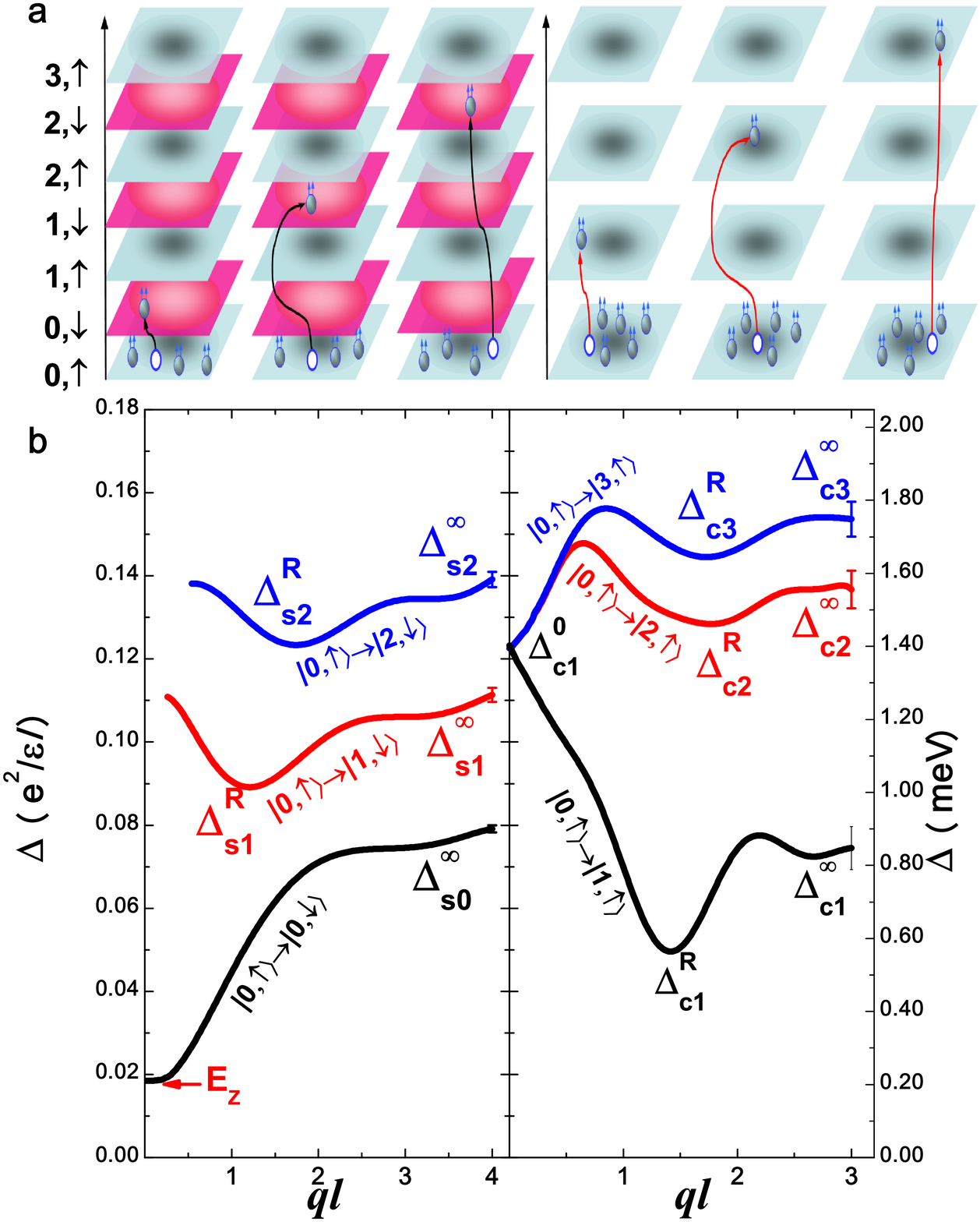}
\caption{Schematic diagram of CF excitons
accompanied by theoretical calculations of their dispersions.
(a) The right panel shows pictorially the SC
excitations $|0,\uparrow\rangle$ $\rightarrow$ $|K,\uparrow\rangle$
across K $\Lambda$ levels.  The left panel shows the
spin-flip modes $|0,\uparrow\rangle$ $\rightarrow$
$|K,\downarrow\rangle$ (b) Calculated dispersions of CF excitons for a
35 nm wide GaAs QW with an electron density of
$5.0\times10^{10}$ cm$^{-2}$. The right (left)
panel shows the dispersions for SC (SF) modes.  The error bar at the end of
each curve represents the typical statistical uncertainty in the
energy determined by Monte Carlo method. Critical points in the
dispersion are labeled.
}
\label{charge}
\end{figure}

\begin{figure}
\includegraphics[width=2.7in,height=2.in]{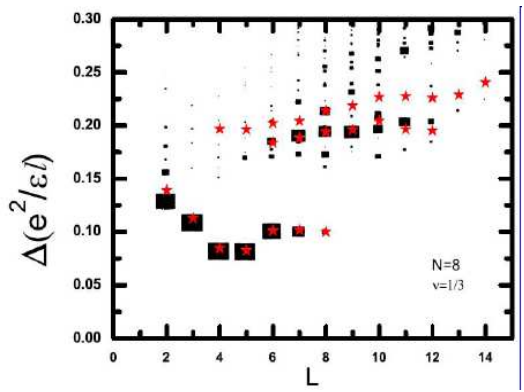}
\caption{Comparison of CF excitons with exact diagonalization results (in spherical geometry) for eight particles at $\nu=1/3$. The red stars show the CF exciton dispersions for the lowest three SC branches for this system as a function of the total orbital angular momentum $L$. The exact spectra are taken from Ref.\cite{song94}. The area of each black rectangle is proportional to the normalized spectral weight under the state; larger spectral weight implies greater intensity in ILS.
The level-1 and level-2 CF excitons closely trace lines
of high spectral weight; it is possible that still higher modes will become identifiable in the exact spectra for larger systems. The other states in the exact spectrum are interpreted as made up of multiple excitons, which are expected to couple less strongly to light.} \label{comp}
\end{figure}

The dispersions of the SC and SF excitons are
obtained by the method of CF diagonalization (without {\em Landau} level
mixing and disorder) \cite{CFDiag1}.
For a more accurate comparison, we have included
here two realistic effects: The finite width
modification of the interaction is incorporated via a self-consistent
local density approximation. We
also allow $\Lambda$ level mixing by considering the
{\em five} lowest energy CF excitons. A combination of these two effects
results in a 20 $\%$ reduction of the energy of the level-2 and
level-3 excitons, and a smaller ($\sim$ 10 $\%$) reduction in the
energy of the level-1 exciton.
Figure \ref{charge} shows the full theoretical dispersions of the CF
exciton branches for SC and SF modes. To avoid clutter,
only the lowest three branches are shown.  The calculations are
performed for 200 (100) particles for SC (SF) modes and
reflect the thermodynamic behavior.  The three dispersion curves
indicated in Fig. \ref{charge}b are assigned as level-0, level-1 and
level-2 for SF modes and level-1, level-2 and level-3 for SC
modes, in order of increasing energy.  The residual interaction between CFs in principle mixes the different ``unperturbed" level-K excitations; however, the modes do not mix significantly at large $ql$, which allows us to continue to use the level-K nomenclature even for mixed modes. Figure \ref{comp} shows a comparison of the CF excitons with exact diagonalization studies on a finite system.

\begin{figure}
\includegraphics[width=2.9in,height=2.2in]{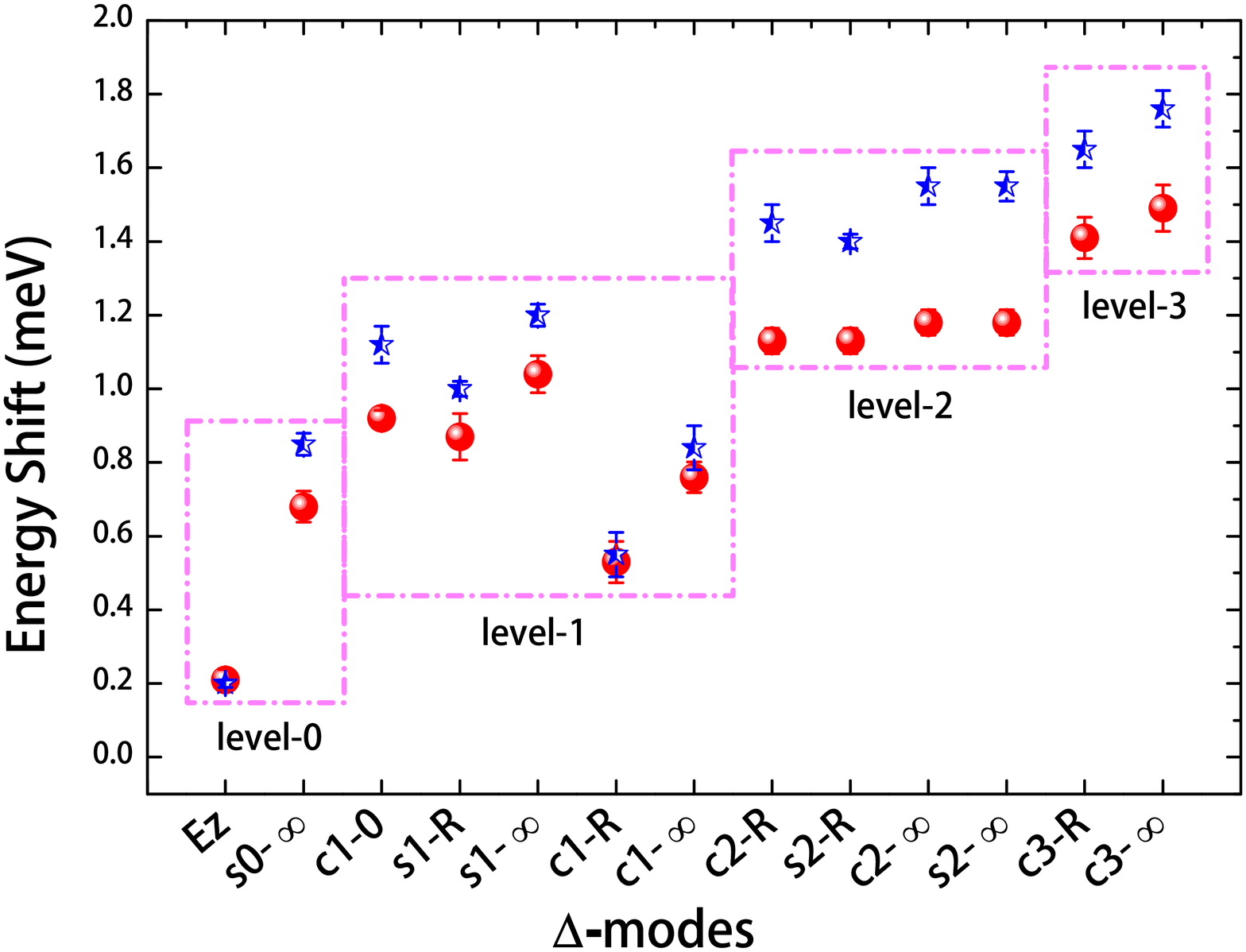}
\caption{ Comparison between experimental energies (from Fig.
\ref{1_3plot}, red circles) with theoretical CF exciton energies
(from Fig. \ref{charge}, blue stars), organized according to the
level of the excitation. The identification of experimental modes is
explained in the text. The discrepancy between theory and experiment,
less than 0.2-0.3 meV, is presumably due to
disorder. Estimated error bars for the experimental values are
shown, unless smaller than the symbol size. } \label{xptvstheory}
\end{figure}

Level-1 SC modes and level-0 SF modes have been
identified in previous experiments
\cite{Hirjibehedin05,Groshaus,Davies97}.
Of interest here are the higher lying
modes. We proceed by sorting the experimental values of the new modes
in ascending order and match them up with theoretical values. The resulting
comparison between theory and experiment is shown in Fig.
\ref{xptvstheory}.
The theoretical results for the energies of level-1
excitons are in excellent agreement with the experimental results.
The only exception is the long wavelength collective mode
$\Delta^0_{c1}$, for which the discrepancy is closer to 35$\%$, but
a $\sim$20$\%$ agreement is achieved when screening of the single
exciton by two-roton excitations is taken into account
\cite{Park00}. This correction, not included in the
calculation shown in Fig. \ref{charge}, is incorporated in Fig.
\ref{xptvstheory}.

It is significant that mode energies predicted by
theory agree to within 0.2-0.3 meV with measured energies, which
translates into a better than 20$\%$ agreement. It should be
stressed that a similar level of deviation between the theoretical
and experimental values of the excitation energies has been found
in the past for other excitations, and attributed to
disorder.  We judge the overall comparison between
theory and experiment to be good, and take it as a strong support of the
identification of the high energy collective modes ranging from
about 1.0 meV to 1.6 meV in terms of transitions of CFs into higher levels.

We note that due to the presence of a large
number of modes, sometimes two or more modes happen to lie at very
nearby energies, and thus may not be resolved in our
experiments. For example, for SC modes, the energy of the level-3
roton overlaps with the small q (q\emph{l} $\sim$ 0.6) critical
point of the level-2 exciton (see Fig. \ref{charge}b). As another
example, the small q (q\emph{l} $\sim$ 0.8) critical point of the
level-3 exciton overlaps in energy with the large wave vector limit
of the level-3 exciton. When encountering such a situation, we have,
for simplicity, arbitrarily assigned one of the possible labels to
the observed mode ($\Delta^R_{c3}$ and $\Delta^\infty_{c3}$,
respectively, for the above two cases). The assignment remains
tentative in such cases, and more sensitive experiments in the
future may reveal further finer structure.

Our work sets the stage for further investigations in other FQH
states in GaAs, and also in other 2D systems, such as
graphene, where the FQH physics is in its infancy
\cite{Andrei09,Bolotin09}. The high energy excitations should also
be accessible to other experimental methods such as
optical absorption\cite{Smet09} and time
domain capacitance spectroscopy\cite{Ashoori10};
these probes are likely to provide important
further insight into the physics discussed above.

\textbf{Acknowledgments} - T.D.R. and A.P. were supported by the
National Science Foundation (NSF) under grants DMR-0352738 and
DMR-0803445; by the Dept. of Energy under grant DE-AIO2-04ER46133;
 and by the Nanoscale Science and Engineering
Initiative of the NSF under award CHE-0641523.
J.K.J. was supported in part by the NSF under
DMR-1005536. The computation was performed at the
Dept. of Theoretical Physics, Indian Assoc. for the
Cultivation of Science.


\begin{thebibliography}{28}
\bibitem[{not()}]{note2}
\bibinfo{note}{Our analysis assumes, as appropriate for weak disorder, that the
  mode wave vector remains a good quantum number and the net effect of disorder
  is to allow coupling to large wave vector excitation modes
  \cite{Marmorkos92,Groshaus}.}
\bibitem[{com()}]{comment4}
\bibinfo{note}{The resonance enhancement of the intensity of excitations in
  ILS experiments depends on the
  frequency of $\omega_L$ and on the energy of the FQH mode. Not
  all modes are visible in a single spectrum and a scan over a range of
  $\omega_L$ energies is necessary to obtain a complete picture.}
\end{thebibliography}

\end{document}